\documentclass[epj,nopacs]{svjour}
\usepackage{graphics}
\usepackage{ulem}
\newcommand{\AmS}{{\protect\the\textfont2
  A\kern-.1667em\lower.5ex\hbox{M}\kern-.125emS}}
\def\ga{\mathrel{\mathchoice {\vcenter{\offinterlineskip\halign{\hfil
$\displaystyle##$\hfil\cr>\cr\sim\cr}}}
{\vcenter{\offinterlineskip\halign{\hfil$\textstyle##$\hfil\cr>\cr\sim\cr}}}
{\vcenter{\offinterlineskip\halign{\hfil$\scriptstyle##$\hfil\cr>\cr\sim\cr}}}
{\vcenter{\offinterlineskip\halign{\hfil$\scriptscriptstyle##$\hfil\cr>\cr
\sim\cr}}}}}

\begin{document}
\hugehead
\title{Pion Freeze-Out Time in Pb+Pb Collisions at 158~A~GeV/c Studied via $\pi^-/\pi^+$ and $K^-/K^+$ Ratios}
\author{
M.M.~Aggarwal\inst{1} \and
Z.~Ahammed\inst{2}  \and
A.L.S.~Angelis\inst{3} * \and
V.~Antonenko\inst{4} \and
V.~Arefiev\inst{5} \and
V.~Astakhov\inst{5} \and
V.~Avdeitchikov\inst{5} \and
T.C.~Awes\inst{6} \and
P.V.K.S.~Baba\inst{7} \and
S.K.~Badyal\inst{7}
% ...
\\ 
\begin{center} (WA98 collaboration) \end{center}
}
\institute{
University of Panjab, Chandigarh 160014, India
\and Variable Energy Cyclotron Centre,  Calcutta 700 064, India
\and University of Geneva, CH-1211 Geneva 4,Switzerland
\and RRC Kurchatov Institute, RU-123182 Moscow, Russia
\and Joint Institute for Nuclear Research, RU-141980 Dubna, Russia
\and Oak Ridge National Laboratory, Oak Ridge, Tennessee 37831-6372, USA
\and University of Jammu, Jammu 180001, India
% ...
}
\date{Received: 15 July 2006}
%
\pagebreak
\newpage
\abstract{
The effect of the final state Coulomb interaction on particles produced
in Pb+Pb collisions at 158~A~GeV/c has been
investigated in the WA98 experiment through the study of
the $\pi^-/\pi^+$ and $K^-/K^+$ ratios 
measured as a function of $m_T$~-~$m_0$.
While the ratio for kaons shows no significant $m_T$ dependence, 
the $\pi^-/\pi^+$ ratio is enhanced at small $m_T$~-~$m_0$ values with an 
enhancement that increases with centrality.
A silicon pad detector located near the target is used to estimate
the contribution of hyperon decays to the $\pi^-/\pi^+$ ratio.
The comparison of results with predictions of the RQMD 
model in which the Coulomb interaction has been incorporated 
allows to place constraints on the time of the pion freeze-out. 
} 
\maketitle

\section{Introduction}

The distributions of negatively and positively charged pions produced
in heavy ion collisions exhibit significant differences at low transverse
kinetic energy  (or transverse mass), $m_T - m_\pi = \sqrt{p_T^2+m_\pi^2}-m_\pi$. 
These differences are evident 
in the behaviour of the ratio of their yields 
$R_\pi$=$\pi^-/\pi^+$. The study of charged particle production in
Au+Au collisions at SIS (1~A~GeV) \cite{SIS} and AGS
(10.8~A GeV)~\cite{AGS_1,AGS_2,Muntz} shows that for
central collisions $R_\pi$ rises at low $m_T$ at both collision energies.
At the same time, for peripheral collisions at the AGS, $R_\pi$ does not
depend on $m_T$. At 1~A~GeV $R_\pi$ is about 2.9 for $m_T-m_\pi$
near zero and its integrated value is $R_\pi$=1.9$\pm$0.1. 
The large integrated value at 1~A~GeV can be explained by isobar
decays and reflects the N/Z asymmetry of the colliding system.
The very sharp rise at low $m_T$ on the other hand points toward 
significant Coulomb interactions between charged pions and the remaining
nuclear charge distribution (acceleration of $\pi^+$ and deceleration
of $\pi^-$).
At 10.8~A~GeV and higher energy the isobar contribution to
$R_\pi$ becomes insignificant and it is expected that mainly the Coulomb
interaction distorts the $R_\pi$ ratio. The NA44 experiment has 
published data on $R_\pi$ measured in central Pb+Pb collisions at
158~A~GeV~\cite{NA44}. A prominent enhancement of $R_\pi$ at
small $m_T$ was observed.
NA44 has also measured $R_\pi$ in S+S and S+Pb collisions at 
200~A~GeV/c, where  the ratio remains constant at
$m_T-m_\pi <$ 0.5~GeV/c$^2$~\cite{NA44}. 
At high energies (AGS and SPS) the contribution from strange particle
decay can significantly affect the measured pion distribution.
The hyperon contribution is influenced by the detector acceptance and the pion vertex
reconstruction properties. The excess of hyperon over anti-hyperon yield
in heavy ion collisions also leads to an enhancement of $R_\pi$ at low
transverse kinetic energy. 

Several dedicated models~\cite{Barz1,Barz2,Heiselberg,Ayala,Scal,RHIC} describe the pion
ratio detected by the NA44 collaboration in Pb+Pb collisions through
the electromagnetic interaction induced by the large amount of charge of
the participating protons. The authors of the dynamical model of Ref.~\cite{Barz1}
argue that
the $R_\pi$ enhancement at small momenta depends mainly on the time of
freeze-out and extract a freeze-out time of about 7~fm/c.
They also predict the corresponding ratio for kaons, $R_K=K^-/K^+$, to be smaller
by a factor of $m_K/m_\pi$. However, this overestimates the NA44 kaon measurement. 
This discrepancy is explained as due to the much larger absorption of 
$K^-$ on nucleons $(K^- + N \rightarrow
\Lambda + \pi)$  than for $K^+$.
For a detector located at mid-rapidity this model predicts a
small reduction of the $R_\pi$ enhancement with respect to the NA44
result measured near rapidity $y_{CMS}=1$ in the center of mass system.
On the other hand, in another model~\cite{RHIC} the 
enhancement of the ratio 
strongly increases towards mid-rapidity, reaching the 
value of 2.5. 
In ref.~\cite{Ayala} it is concluded that the influence of the
Coulomb force should be computed within an event simulator which takes
into account a more accurate description of the space-time evolution of
the collision.

Initial results from the SPS experiment WA98 on the measurement of the $\pi^-/\pi^+$ ratio in
central Pb+Pb collisions  were reported
in Ref.~\cite{Fabrice}. The enhancement of $R_{\pi}$ at $m_T-m_\pi < $50~MeV/c$^2$
measured with the two independent tracking arms of WA98 located nearer
to mid-rapidity than the NA44 spectrometer was reported to be significantly
smaller than that measured by NA44~\cite{NA44}.

In this paper we present the $\pi^-/\pi^+$  ratio as a function of
centrality for 158 A GeV/c Pb+Pb collisions.
The background pion contribution from hyperon decays is determined and removed 
by use of vertex information from a silicon detector located near to the target
to obtain the ratio of pions that originate in the target.
The yields of electrons and positrons are analysed at low momentum where
they can be clearly identified through time-of-flight. Their yields provide a check
of the normalization of the opposite sign yields at the lowest transverse momenta.
They also confirm  the validity
of estimates of the electron and positron contaminations
to the measured pion distributions at higher momenta. 
The $K^-/K^+$ratio is also presented for central collisions.
The results  are
compared with predictions of the RQMD~2.4 model~\cite{RQMD}   
in which final state Coulomb interactions have been added. 
Within the uncertainties of the hyperon decay contribution 
the results are found to be compatible with
predictions of the RQMD model in which the average pion freeze-out time is 15 fm/c for
central collisions.
Possible modifications of the final state predicted by RQMD, necessary to  improve
agreement with the measured $\pi^-/\pi^+$ ratios, are investigated and
discussed.

\begin{figure*}
\rotatebox{270}{
 \resizebox{0.7\textwidth}{!}{
  \includegraphics{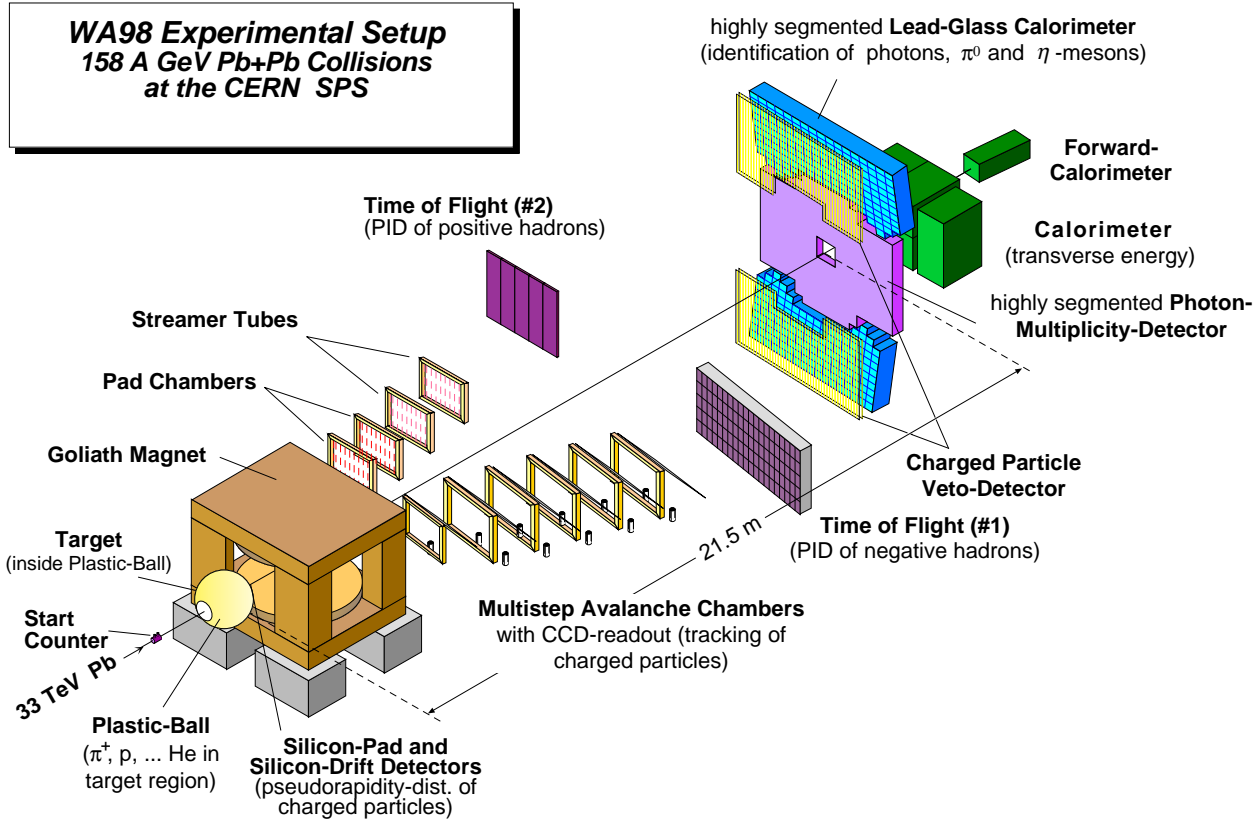}}}
\vspace*{0cm}
\caption{The WA98 experimental setup.}
\label{fig:wa98setup}
\end{figure*}

\section{Experimental Setup and Data Analysis}

The CERN SPS experiment WA98~\cite{misc:wa98:proposal:91} combined photon
and hadron spectrometers with other large acceptance detectors that measured
a number of global variables on an event-by-event basis.
Fig.~\ref{fig:wa98setup} shows the experimental layout
during the 158~A~GeV $^{208}$Pb beam run period 
in 1996 with a 0.239~g/cm$^2$ Pb target.

The Mid-Rapidity Calorimeter, MIRAC~\cite{mirac}, measured the collision
transverse energy $(E_T)$ in the pseudorapidity interval $3.5 < \eta < 5.5$.
The minimum bias trigger required $E_T\ga$~5~GeV 
and a valid signal from
the beam counters. The measured $d\sigma/dE_T$ distribution is used for the
calculation of the collision centrality. The minimum bias cross section for the
run period used in this analysis is $\sigma_{mb}$=6451~mb with an overall
systematic error of less than 10$\%$.

Two thin circular silicon wafer detectors were positioned at small distance 
downstream from the target, where charged particle trajectories were not
affected by the magnetic field. A Silicon Pad Multiplicity Detector,
SPMD~\cite{spmd}, was located at a downstream distance of 32.85~cm. It was segmented
into 184 azimuthal sectors and 22 pseudorapidity rings in the range
$2.35 < \eta <3.75$, and maintained pad occupancy below 20$\%$.
The SPMD measured the energy deposited in each pad.
A Silicon Drift Detector, SDD~\cite{sdd}, was  located at a distances of
12.5~cm from the target. Its position resolution was 25 and 35~$\mu$m in the
azimuthal and radial coordinates, respectively.
The acceptance of the two detectors overlapped in the pseudorapidity interval
$2.35 < \eta <3.4$. The two detectors were used for precise reconstruction of
the vertex position of the Pb+Pb primary collision.
For that purpose straight lines going through each pair of hits in the SDD
and SPMD were projected onto the target plane.
The one dimensional spatial resolution was found to be better than 
0.3~mm perpendicular to the beam line.

The WA98 experiment comprised two charged particle spectrometer arms located on
the right (first arm) 
and the left (second arm, not used in the present analysis) 
facing downstream from the target, beyond a dipole 
magnet (Goliath) with 1.6~Tm bending power in the horizontal plane.
The first tracking arm consisted of six Multistep Avalanche Chambers, MSAC, 
read out by CCD cameras equipped with two image intensifiers~\cite{MSAC}.
Each pixel of a CCD viewed a 3.1$\times$3.1~mm$^2$ area of a chamber.
In addition, a 4$\times$1.9~m$^2$ Time-of-Flight wall positioned behind
the chambers at a distance of 16.5~m from the target allowed for particle
identification with a time resolution better than 120~ps. The TOF detector
consisted of 480 scintillator counters with area 3.3$\times$48.5~cm$^2$ 
and thickness 2~cm arranged in four rows~\cite{RTOF}. Each counter was
equipped with two PMTs, one at each end. The position of a hit along a
scintillator bar was evaluated using two methods: the first used the
time difference of signals obtained from the two ends of the scintillator
bar, and the second used the ratio of their amplitudes.
Both methods provide a spatial resolution of the order of 2.5~cm.

\begin{figure}[!h]
  \resizebox{0.5\textwidth}{!} { \includegraphics{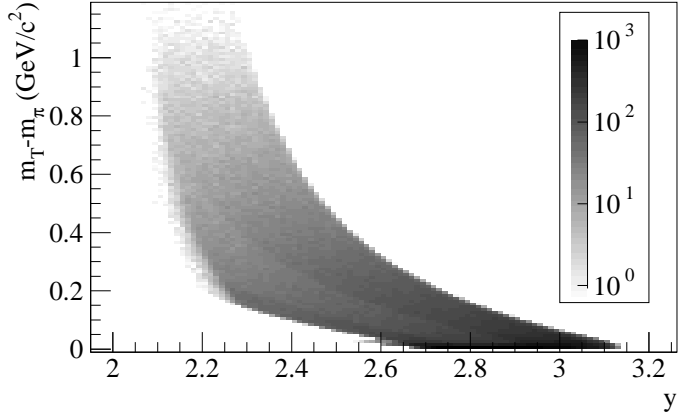} }
  \vspace*{0.cm}
  \caption{Acceptance of the first tracking arm
            in the $(m_T-m_\pi,y)$ plane for pions.}
\label{fig:accept}
\end{figure}

Tracks were selected which traverse all six MSAC chambers, have detected
hits in at least four of them, and were associated with a TOF hit within a
6$\times$6~cm$^2$ window. 
An additional momentum dependent time of flight cut was used to exclude
misidentified pion, kaon, and proton tracks at a level better than 1\%.
The $(m_T-m_\pi)$~--rapidity acceptance for pions generated by Monte Carlo
is shown in Fig.~\ref{fig:accept}. The center of mass system rapidity
is 2.9. The momentum resolution of the spectrometer was 
$\Delta p / p = 0.005$ at p = 1.5~GeV/c.

Positive and negative particles were detected in two sets of runs with
opposite direction of the magnetic field. Data samples consisting of
0.58$\times$10$^6$ $\pi^-$ and 0.22$\times$10$^6$ $\pi^+$ tracks satisfying
all quality requirements were used for the ratio measurement.

Efficiency matrices were measured for a grid of 8$\times$8 cells across
each MSAC chamber and for several run periods. 
The overall tracking arm efficiency
measured for negative particles was 2\% less than that for positive 
particles due to chamber instabilities,
however, there was no significant difference in the $m_T$ dependence
of the efficiencies.
The detailed analysis of the $\pi^-$ and $K^-$ measurement 
in the first tracking arm has been presented 
in a separate paper~\cite{Sandor_spectra}.

\section{Method of Simulation}

Simulations have been used in this analysis to estimate the different contributions 
to the measured $R_\pi$ and $R_K$ ratios, and to compare
with the measured results. 
A detailed description of the primary collisions, of the Coulomb final state
interactions, and of the particle propagation through the
experimental setup has been simulated.

The primary Pb+Pb collisions were simulated with the RQMD~2.4 model~\cite{RQMD}   event
generator which reproduces well the light strange baryon yield measured by the
WA97 experiment~\cite{WA97RQMD}, but does not include Coulomb
interactions. In RQMD the particles are treated
semiclassically over the complete evolution of the system. This 
allows the calculation of the Coulomb forces between each particle pair
as an "afterburner" applied to the RQMD output. 
The electromagnetic interaction of a particle $i$ with another charged
particle $j$ was calculated as described in Ref.~\cite{RelCoulomb}
by 
\begin{eqnarray}
m_i \frac{du^\mu_i}{d\tau} = \sum_{j \ne i} \frac{ e_i }{c} 
                                        F_{ij}^{\mu\nu} u^\nu_i,
\end{eqnarray}
where $j$ runs over all charged particles of the system.
Here particle $i$ has mass $m_i$, charge $e_i$ and 4-velocity
$u_i=(1,\mathbf{v})/\sqrt{1-\mathbf{v}^2}$. 
The tensor $F_{ij}^{\mu\nu}$ is given by 
\begin{eqnarray}
F_{ij}^{\mu\nu} = \frac{ e_j }{c} \frac{X^\mu u^\nu_j - X^\nu u^\mu_j}
  {(\frac{1}{c^2} (u_j^\lambda X^\lambda)^2 - X_\lambda X^\lambda)^{3/2} }
\end{eqnarray}
where $X^\lambda$ is the relative 4-distance between particles $i$ and $j$.
The forces are calculated at each time step of the system evolution and 
the positions and momenta 
of the particles are updated with the effect of the Coulomb forces 
only for those particles which have undergone their last collision.
RQMD events with Coulomb interactions added were used as input to the detector simulation. 
Delta electrons produced by Pb ions in the target
material before the collision were also simulated through the detector response.

The GEANT~3.21 program~\cite{GEANT} was used for the detector simulation.
The target, beam line, Goliath magnet, SDD and SPMD detectors, and 
tracking arm have been described in detail. 
For the track reconstruction the same procedure 
as used in the treatment of real data was employed. 
For every hit in the tracking arm all information on
the particle and all its predecessors was stored for further analysis so that the
history of every track could be traced back to the moment of the primary
collision.

\section{Experimental Results}

\subsection{Centrality selection}

The $R_\pi$ ratio was studied for six centrality intervals defined by the
transverse energy $E_T$ measured with the MIRAC.
The value of $b$ may be estimated from the measured $E_T$ according to the equation
\begin{eqnarray}
\pi b^2 = \int\limits_{E_T}^\infty (d\sigma/dE_T) dE_T,
\end{eqnarray}
where $d\sigma/dE_T$ is the experimentally measured distribution. Systematic
comparisons of $d\sigma/dE_T$ with predictions of VENUS~4.12~\cite{VENUS} and
extraction of the number of participating nucleons or number of binary
collisions are given in Ref.~\cite{etscaling}.
Table ~\ref{tab:table1} lists the centrality intervals used in this analysis.

\begin{table}[h]
\label{tab:table1}
\begin{tabular}{|l|l|l|l|l|}   \hline
Centrality  &   &  & Impact & Number \\
interval& $E_T$ & $~\sigma/~\sigma_{mb}$  & parameter & of parti-\\
number  & (GeV) & (\%) & (fm) & cipants \\ \hline
1 & 34.5-130 & 36.8-70.2 & 8.6-12.0 & 34-136   \\ \hline
2 & 130-180  & 28.0-36.8 & 7.5-8.6 & 136-179   \\ \hline
3 & 180-240  & 19.6-28.0 & 6.3-7.5 & 179-231   \\ \hline
4 & 240-325  & 10.1-19.6 & 4.5-6.3 & 231-306   \\ \hline
5 & 325-400  & 3.58-10.1 & 2.7-4.5 & 306-364   \\ \hline
6 & $>$400   & $<$3.58    & $<$2.7 & $>$364   \\ \hline
\end{tabular}
\caption[]{Centrality intervals defined according to the amount of $E_T$
measured in MIRAC. The number of participants was calculated with 
VENUS~4.12.}
\end{table}

\subsection{Particle ratios}

\begin{figure}[!h]
\resizebox{.5\textwidth}{!} { \includegraphics{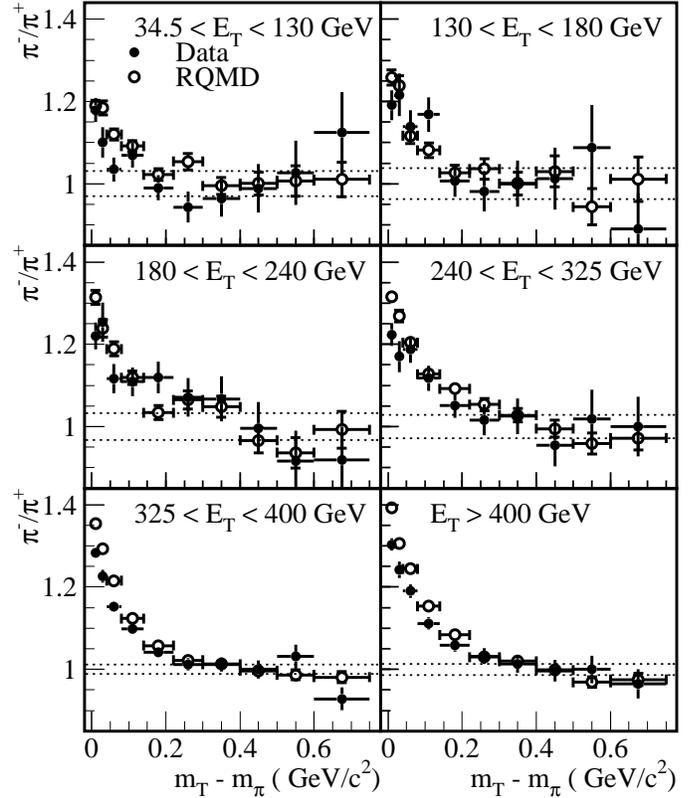} }
  \vspace*{0cm}
  \caption{ The ratio $R_\pi$ vs $m_T-m_\pi$ for
  several Pb+Pb collision centrality intervals defined by $E_T$ measured in the MIRAC.
  Data (filled circles) compared with predictions
  from RQMD including the Coulomb interaction (open circles).
  The ratios are arbitrarily normalized to 1.0 in the region
  $0.3 <m_T-m_\pi < 0.76$~GeV/c$^2$.
  The statistical errors are shown by the error bars on the points.
 The dotted lines indicate the statistical error on the
  normalization of the data. The RQMD normalization errors are about half as large.}
\label{fig:arm1centr}
\end{figure}

The two tracking arms of WA98 provide two independent 
measurements of the pion ratio and have been shown to be consistent~\cite{Fabrice}. 
This study focuses on an analysis
of the data obtained by the first tracking arm, whose acceptance for
pions is shifted by about 0.5 of a unit of rapidity closer to the target
fragmentation region relative to the acceptance of the second arm.
This provides more favourable conditions for $\pi/K$ separation in the
region of large transverse mass where the normalization of the ratios is performed.
At low $p_T$ it allows better rejection of electrons through time-of-flight
and a significantly larger number of tracks pass through the active area
of the SPMD than is the case for the second arm.
Fig.~\ref{fig:arm1centr} shows the $R_\pi$ ratio normalized at transverse
kinetic energy between 0.3 and 0.76 GeV/c$^2$ for the six centrality selections
of Table~\ref{tab:table1}. It should be noted that the Coulomb interaction effect 
on the ratio is strongly reduced and the contribution from hyperon decays
is absent in the normalization region.

\begin{figure}[!h]
  \resizebox{0.5\textwidth}{!} { \includegraphics{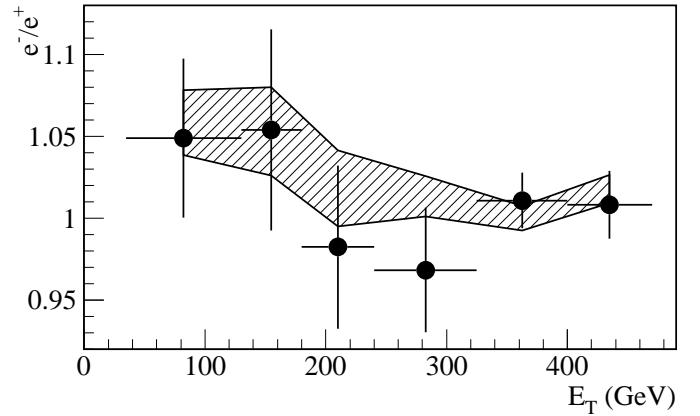} }
  \vspace*{0.cm}
  \caption{MIRAC $E_T$ dependence of the $e^-/e^+$ ratio.
    The normalization is the same as for the $\pi^-/\pi^+$ ratio.
    Data are shown by filled circles.
    The hatched band indicates the statistical uncertainty 
    (Mean$\pm$RMS) of the RQMD prediction.}
\label{fig:EleNorm}
\end{figure}

The $e^-/e^+$ ratio can be used as a check of the absolute normalization
of the $R_\pi$ ratio.
For $0.8 < p < 1.2$~GeV/c electron and positron tracks
are well identified through time-of-flight.
As imposed by the tracking arm acceptance these tracks have
transverse momenta below 50~MeV/c.
Since both electrons and positrons originate
mainly from photon conversions in the target material 
their yields should be identical,
except for a small additional electron contribution from the production
of delta electrons.
These purely electromagnetic processes
are well understood, so that the ratio $e^-/e^+$
is expected to be reproduced in simulation
with an accuracy of order $1\%$.
The $e^-/e^+$ ratio is presented in Fig.~\ref{fig:EleNorm} 
as a function of centrality and compared to predictions.
For each centrality the $e^-/e^+$ ratio is normalized
with the same normalization used for the corresponding $\pi^-/\pi^+$ ratio
in Fig.~\ref{fig:arm1centr}.
Any difference between data and simulation is a measure
of the systematic error in the normalization of the
measured $R_\pi$ distributions of Fig.~\ref{fig:arm1centr}.
Summed over all centrality bins
the difference between data and simulation
of the $e^-/e^+$ ratio is -\,0.8$\pm$1.2$\%$, consistent
with the expected accuracy of the comparison.
This result indicates that the systematic error 
on the absolute normalization of $R_\pi$ introduced by 
normalization of the ratio in the region 
$0.3< m_{T}-m_{\pi} <0.76$~GeV/c$^2$ is small, and that 
the normalization error is dominated by the statistical error
of the data used for the normalization (shown by the dotted
lines in Fig.~\ref{fig:arm1centr}).

\begin{figure}[!h]
  \resizebox{0.5\textwidth}{!} { \includegraphics{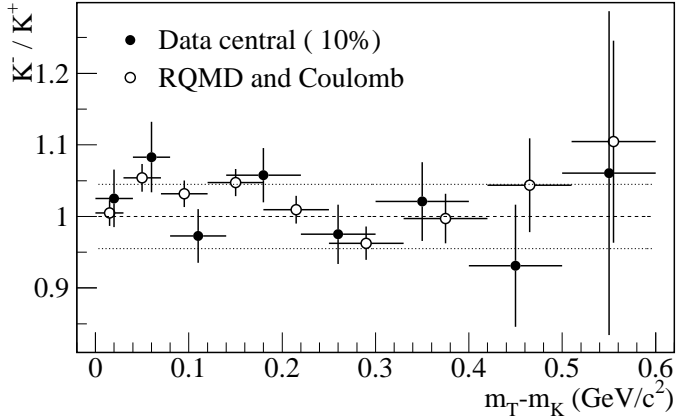} }
  \vspace*{0cm}
  \caption{The kaon ratio for the 10\% most central Pb+Pb collisions.
  Data (filled circles) are compared with predictions
  from the RQMD model with Coulomb interactions included (open circles).
  The ratios are arbitrarily normalized to 1.0 in the region
  $0.3 <m_T-m_K <0.6$~GeV/c$^2$.
   Errors are as described for Fig.~\ref{fig:arm1centr}.}
\label{fig:kaonratio}
\end{figure}

The ratio for kaons $R_K$ measured under the same 
experimental conditions is
presented in Fig.~\ref{fig:kaonratio} for the 10\% most central
collisions. The kaon ratio shows no significant dependence on
transverse kinetic energy. The lack of enhancement at low transverse mass
is consistent with RQMD simulations with Coulomb
final state interactions.
The acceptance for the lowest transverse mass is located near kaon 
rapidity $y$=1.8 in the laboratory system. For $m_T-m_K > 0.2$~GeV/c$^2$ 
the average rapidity is about $y$=2.15.

\subsection{Hyperon contribution}

Pions from hyperon decays have a much steeper transverse momentum
spectrum than directly produced pions.  Since strange hyperons are more
abundantly produced than anti-strange hyperons, the $\pi^-/\pi^+$ ratio 
will be enhanced at low $m_T$ due to this difference. In order to study the
Coulomb effect on the directly produced pions, it is necessary to estimate
and remove the hyperon contributions.  

The Silicon Pad Multiplicity Detector has been used to estimate the yield of identified tracks that 
do not track back to the target vertex, and thus may be
attributed to products of strange or anti-strange hadron
decays.  All pads of the SPMD that overlap with a 4$\times$4~mm$^2$ 
window, centered at the point where the extrapolation of 
the track towards the target intersects the SPMD plane, were selected. 
The summed energy loss in the pad window is denoted as $S$, and given in units
of the energy loss of a minimum ionizing particle in
the silicon wafer of the SPMD. $S$ therefore gives an estimate of the
number of minimum ionizing particles traversing the SPMD within the pad window. 
Simulation indicates that the mean value over a large set of tracks for pion tracks
originating in the target is $\left\langle S \right\rangle$ = 0.89, while
for tracks from hyperon and $K^0_S$ decays 
$\left\langle S \right\rangle$=0.17 and 0.20, respectively.
This is because tracks that originate from hyperon or $K^0_S$ decays 
either traverse the SPMD far from the pad indicated by the extrapolation, 
or do not traverse it at all (produced downstream of the SPMD).
Thus,  tracks from hyperon decays 
will be suppressed by roughly a factor of $0.89/0.17 \approx 5$ 
with the SPMD hit requirement.

In the case of high pad occupancy the energy loss $S_1$ measured in the
SPMD window associated with the track contains a significant contribution $S_2$
from
spurious tracks. The value of $S_2$ is estimated by the measured energy loss
in a window of the same size centered at a location far from the extrapolated track
location (in particular, at a location rotated by $12^\circ$ in azimuth).
The RQMD simulation indicates that  $\left\langle S' \right\rangle 
= \left\langle S_1-S_2\right\rangle$
provides a good estimate of $\left\langle S \right\rangle$ for each kind of
track.

\begin{figure}[!h]
  \resizebox{0.5\textwidth}{!} { \includegraphics{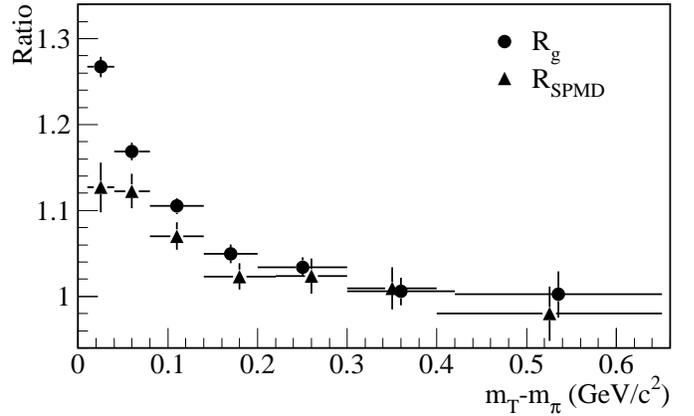} }
  \vspace*{0cm}
  \caption{$R_{SPMD}$ (triangles) and $R_g$ (circles) for
   the $10\%$ most central events.
   See text for definitions of $R_{SPMD}$ and $R_g$.
   Both distributions are normalized to 1.0 for $R_g$ at 
   $m_T-m_\pi > 0.3$~GeV/c$^2$.}
\label{fig:gspmd}
\end{figure}

For tracks whose extrapolation traverses the SPMD the
quantities $R_g=\pi^-/\pi^+$ and $R_{SPMD}=S'^-/S'^+$ are defined, 
where $S'^-$ and
$S'^+$ are sums of $S_1-S_2$ values for negative and positive pion
tracks identified by the tracking arm. In the case of equal hyperon fractions
of the pion yield for both charges one expects $R_{SPMD} \approx R_g$.
The behaviour of $R_g$ and
$R_{SPMD}$ as a function of transverse kinetic energy for 
the 10$\%$ most central collisions is shown in Fig.~\ref{fig:gspmd}.
Both distributions have the same normalization chosen to give $R_g=1$ at 
$m_T-m_\pi > 0.3$~GeV/c$^2$. 

\begin{figure}[!h]
  \resizebox{0.5\textwidth}{!} { \includegraphics{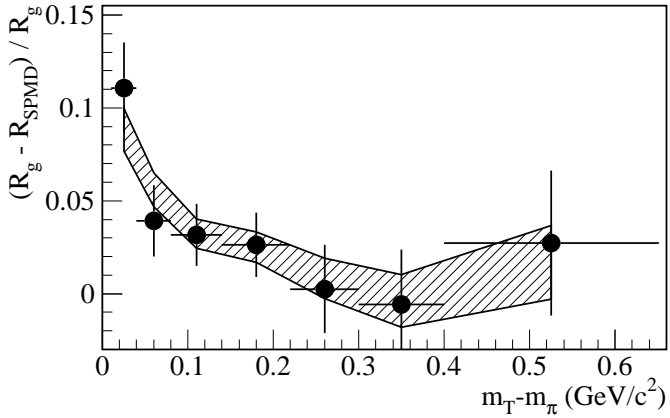} }
  \vspace*{0cm}
  \caption{ $(R_{SPMD}-R_g)/R_g$ for the $10\%$  most central events.  
    Data are shown by filled circles. See text for definitions of $R_{SPMD}$ and $R_g$.
    The hatched band indicates the statistical uncertainty (Mean$\pm$RMS)  of the RQMD prediction.}
\label{fig:hyp_shape}
\end{figure}

\begin{figure}[!h]
  \resizebox{0.5\textwidth}{!} { \includegraphics{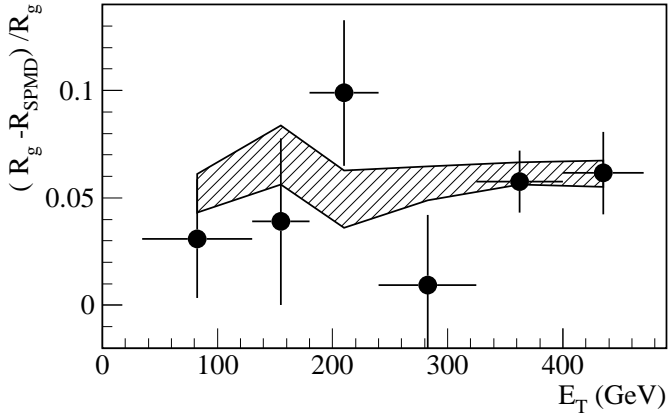} }
  \vspace*{0cm}
  \caption{Centrality dependence of
    $(R_{SPMD}-R_g)/R_g$ for $m_T-m_\pi <$~140~MeV/c$^2$.
    Data are shown by filled circles.
    The hatched band indicates the statistical uncertainty (Mean$\pm$RMS) of the RQMD prediction.}
\label{fig:hyp_yield}
\end{figure}

Since the contribution to $S$ from hyperon decay tracks is smaller
than for tracks originating in the target, a deviation of $R_{SPMD}$
from $R_g$ is an indication of different hyperon fractions in the two
charge samples.  The observed
lower value of $R_{SPMD}$ in comparison to $R_g$ for small $m_T$ 
indicates a larger fraction of tracks from hyperon decays in 
the measured $\pi^-$ distribution relative to the $\pi^+$ distribution.
To quantify this difference the quantity $(R_g-R_{SPMD})/R_g$ 
is extracted and compared with RQMD simulation 
as a function of $m_T-m_\pi$ in Fig.~\ref{fig:hyp_shape} and 
as a function of collision centrality in Fig.~\ref{fig:hyp_yield} 
for $m_T-m_\pi <$~140~MeV/c$^2$, where the excess becomes significant.
Within errors it is seen that the RQMD simulation provides a good
description of the contribution of strange baryon decays 
to the pion ratio. 

\begin{figure}[!h]
\resizebox{.5\textwidth}{!} { \includegraphics{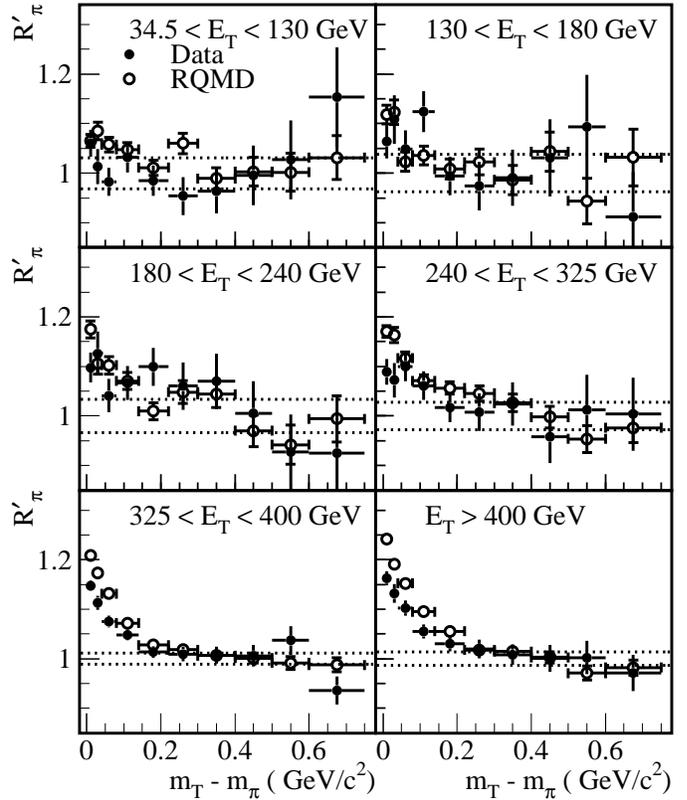} }
  \vspace*{0cm}
  \caption{ The corrected charged pion ratio $R^\prime_\pi=\pi^-/\pi^+$ for 
  pions originating from the target, after removal of the hyperon decay
  and electron contamination contributions.
  Data (filled circles) are compared with predictions
  from RQMD including the Coulomb interaction (open circles).  
  The ratios are arbitrarily normalized to 1.0 in the region
  $0.3 <m_T-m_\pi < 0.76$~GeV/c$^2$.
   Errors are as described for Fig.~\ref{fig:arm1centr}.
}
\label{fig:arm1_centr_nohyp}
\end{figure}

The yield distributions of light strange and anti-strange baryons have been reported by the WA97
collaboration [22] and also shown to be successfully reproduced by RQMD calculations. 
Based on this fact, and on the comparisons of Figs.~\ref{fig:hyp_shape} and ~\ref{fig:hyp_yield},
the RQMD simulation results have been used to correct the
measured pion ratios $R_{\pi}$ shown in Fig.~\ref{fig:arm1centr} for the
hyperon decay contributions to obtain the corrected ratios $R^\prime_\pi$ 
of pions originating in the target. 
For this purpose, for each of the $\pi^-$ and $\pi^+$ yields, 
the fraction of the pion yield from strange hadron decays has
been estimated from RQMD simulation and removed from the total yield.
Fig.~\ref{fig:arm1_centr_nohyp} shows the corrected pion ratios 
$R^\prime_\pi=\pi^-/\pi^+$.

At low momenta the $e^-$ and $e^+$ tracks have been separated from pions
through time-of-flight and their abundances are found to be in agreement
with simulation. 
At high momenta $e^-$ and $e^+$ tracks cannot be separated from pions
and their relative yields have been estimated from simulation. 
The ratio $R^\prime_\pi$ of pions originating from the target 
has also been corrected for $e^-$ and $e^+$
misidentified as pions at high momenta. 
According to simulation this $e^-$ and $e^+$ contamination
weakens the enhancement of
the pion ratio by about $2\%$ for the lowest $m_T$. 
The corrected ratios $R^\prime_\pi$ shown in
Fig.~\ref{fig:arm1_centr_nohyp} can be used for comparison 
of the $\pi^-/\pi^+$ ratio with predictions from models 
including the decay of the $\eta$ and other short lived resonances.

\begin{figure}[!h]
  \resizebox{0.5\textwidth}{!} 
    { \includegraphics{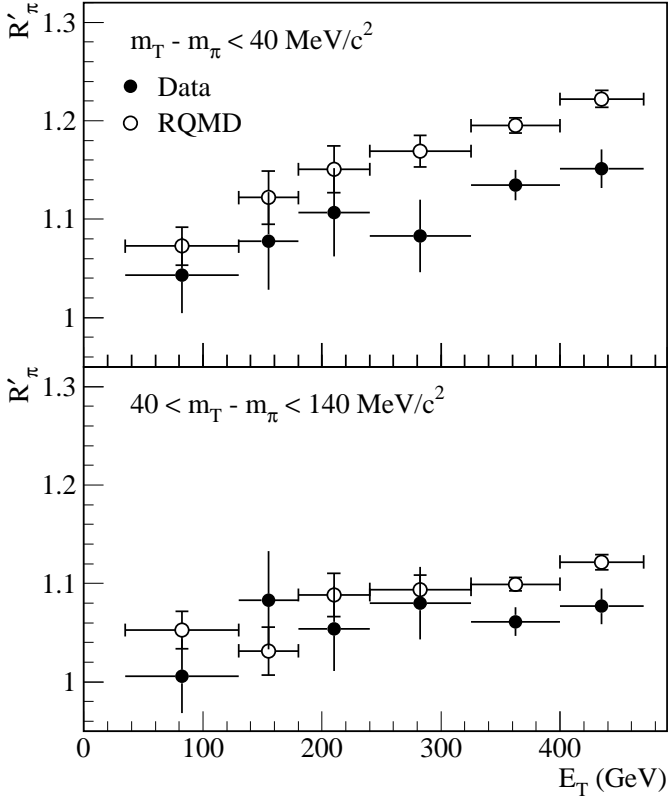} }
  \vspace*{0cm}
  \caption{Centrality dependence of the 
   corrected charged pion ratio for 
   pions originating from the target for $m_T-m_\pi < 40$~MeV/c$^2$
   and $40 < m_T-m_\pi < 140$~MeV/c$^2$.
   Data  (filled circles) are compared with predictions
   from RQMD calculations (open circles).}
\label{fig:arm1_mt40_nohyp}
\end{figure}

In Fig.~\ref{fig:arm1_mt40_nohyp} 
the dependence of the corrected pion ratio $R^\prime_\pi$ 
on the collision centrality is shown for two intervals 
of pion transverse kinetic energy: below 40 MeV/c$^2$
and between 40 and 140 MeV/c$^2$. 
The error due to normalization is included in the
error bars. Predictions from RQMD calculations with
Coulomb interactions included are also shown. Both the measured results
and the RQMD predictions show a smooth increase of the pion ratio 
with increasing centrality. As expected, the ratio tends to 
a value of unity for peripheral collisions. However,  RQMD
is seen to overpredict the measured pion ratio with a 
discrepancy that grows with decreasing transverse mass.

\section{Modified Coulomb Calculations}
Several parameters of the collision, 
such as the total participant charge at central rapidity,
its transverse expansion velocity, and the freeze-out time,
affect the $\pi^-/\pi^+$ ratio~\cite{Barz1}. 
In order to investigate the sensitivity of the pion ratio to the collision parameters,
the Coulomb calculations were repeated under alternative scenarios 
following minor modifications of the RQMD 
output. For this study RQMD events from 
central Pb+Pb collisions with impact parameter $b < $4.5~fm were used.

\begin{figure}[!h] 
\resizebox{0.5\textwidth}{!}{\includegraphics{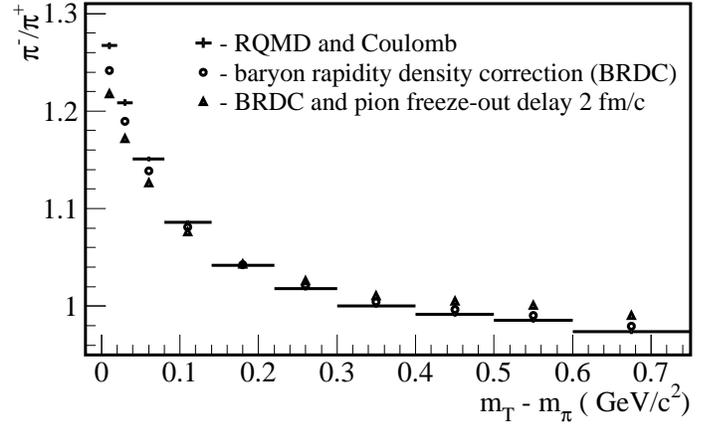}} 
\caption{The pion yield ratio $\pi^{-}/\pi^{+}$ at mid-rapidity  ($1<|y_{CMS}|$). 
Predictions from RQMD with final state Coulomb interaction unmodified (crosses),
 with BRDC (circles), and with BRDC and pion freeze-out delayed an additional 
 2~fm/c (triangles).}
\label{fig:coulbyc2fm}
\end{figure}

In the first modification, the RQMD baryon rapidity density was 
adjusted.
Since the value  of $dN/dy$ for protons around mid-rapidity
for central Pb+Pb collisions as measured by the NA44~\cite{midrap_p} and
NA49~\cite{na49midrap_p} experiments is lower than the 
RQMD prediction by about 30\%, the RQMD Coulomb interaction from baryons
should be reduced accordingly.
The corresponding reduction of the baryon Coulomb source
was implemented in the calculation 
by weighting the Coulomb charge of all baryons during all stages of the collision 
by the ratio of 
measured to predicted proton $dN/dy$ distributions.
This baryon rapidity density correction (BRDC) 
reduces the ratio at the lowest $m_T-m_\pi$ bin by 3\% 
(Fig.~\ref{fig:coulbyc2fm}).
Since this correction is based on a discrepancy between RQMD and 
experimental measurements,
further calculations use this modified baryon
Coulomb charge rapidity density.

To estimate the sensitivity of the pion ratio to the freeze-out time the application
of the Coulomb interaction was simply delayed.
The mean time of pion freeze-out at $|y_{CMS}|<1$ in the central Pb+Pb RQMD events
is 15~fm/c.
For demonstration, a  2~fm/c delay was added during which time the particles move freely
after freeze-out on their final RQMD trajectories. After propagating freely for 2~fm/c the 
pions are then subjected to the Coulomb interactions.
At $m_T-m_\pi < $40~MeV/c$^2$ the resulting ratio 
shown in Fig.~\ref{fig:coulbyc2fm} by triangles 
is reduced with respect to the unmodified RQMD result by 5\%.
This total reduction of the ratio is similar in magnitude to the difference between the data 
and the unmodified RQMD simulation results 
of Fig.~\ref{fig:arm1_mt40_nohyp}.  Alternatively, increased collective flow
in RQMD would result in a larger freeze-out volume and a reduced Coulomb
interaction, with a similarly improved agreement with the measurements.

\begin{figure}[!h]
  \resizebox{0.5\textwidth}{!} { \includegraphics{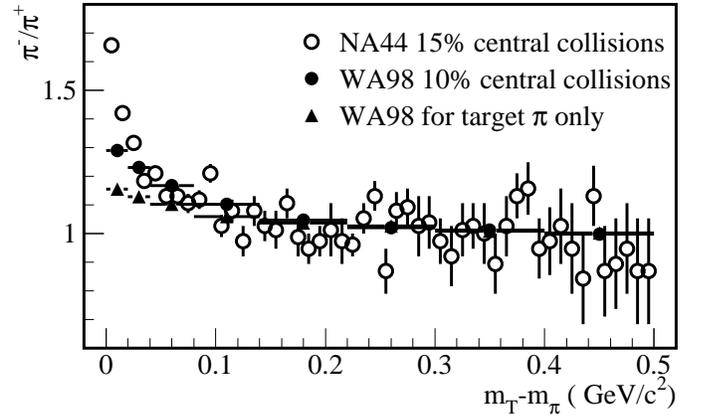} }
  \vspace*{0cm}
  \caption{Comparison of NA44 data (open circles) and
   WA98 data (filled circles) for the 10$\%$ most central collisions.
   The ratio for pions originating in the target is also shown 
   (triangles).}
\label{fig:na44}
\end{figure}

In Fig.~\ref{fig:na44} the pion ratio measured in 
this experiment
for the 10\% most central Pb+Pb collisions is compared with that reported by the NA44
collaboration~\cite{NA44}. 
Neither RQMD calculations, 
nor the previously mentioned models~\cite{Barz1,RHIC}, 
predict the observed difference in the two measurements. 
This could indicate an earlier
pion freeze-out in the center of mass
rapidity region near $y$=1 (NA44) relative to that near $y$=0 (WA98).

The measured $K^-/K^+$ ratio (Fig.~\ref{fig:kaonratio}) does not show any significant $m_T$
dependence. This is consistent with the data of the NA44 collaboration and with
RQMD predictions. The kaon ratio enhancement induced through the Coulomb
interaction is expected to be less than that for pions by a factor equal to
their mass ratio $m_\pi/m_K$~\cite{Barz2}. This relation is satisfied with
respect to the pion ratios presented here, but it is not for
the NA44 data~\cite{NA44}.
Therefore, while the Coulomb effect is consistent with the
WA98 kaon result, an additional mechanism, such as 
different absorption of $K^+$ and $K^-$, is necessary to explain the 
NA44 result.

\section{Discussion and Conclusions}

The  $\pi^-/\pi^+$ ratio has been
measured as a function of transverse mass and centrality in Pb+Pb collisions
at 158 A GeV/c.
The hyperon decay contribution has been deduced and shown to result 
in an increase of the pion ratio at low $m_T$ that is responsible for about half 
of the enhancement of $\pi^-$ relative to $\pi^+$
in central collisions, 
and an even larger fraction of the enhancement in peripheral collisions.
The RQMD model was shown to agree with the measured hyperon decay contributions.
The hyperon decay contributions have been removed from the measured
ratios to extract the ratios for directly emitted pions.

The low $m_T$ enhancement of  the $\pi^-/\pi^+$ 
ratio, after removal of the hyperon decay contribution,
increases with collision centrality and tends to zero
for the most peripheral events (see Fig.~\ref{fig:arm1_centr_nohyp}). 
This observation is consistent with the hypothesis 
that the enhancement is due to the 
Coulomb interaction induced by the net positive charge of the
participant protons. The same behaviour is observed in the framework of
RQMD model simulations that include final state Coulomb interactions.
However,  the RQMD simulation overpredicts the observed enhancement in the ratio 
by a factor of about 1.5. 

Investigations of the RQMD model predictions 
suggest that rather small modifications of the properties 
of the participant fireball are necessary to obtain good 
agreement with the data. For example, good agreement 
can be achieved after reduction of the baryon rapidity charge density predicted by RQMD 
according to the NA44 and NA49 proton measurements, together with a small increase of 
the mean pion freeze-out time (such as an increase from an average of 15 fm/c to 17 fm/c).
According to RQMD predictions, the transverse flow velocities 
of heavy particles are less than those of pions~\cite{Sorge_flow}.
A slight increase of the baryon transverse flow velocity
would have a similar effect as the delay of the freeze-out time.
In both cases, the system disperses over a larger volume before freeze-out, 
which reduces the Coulomb field and, therefore, 
the $\pi^-/\pi^+$ ratio. 
Based on comparisons with the RQMD model calculations,
the results suggest a relatively large value 
of the mean freeze-out time for pions ($\sim$15~fm/c) 
in central Pb+Pb collisions.

\vspace{5 mm}
\hspace{-5.5 mm}
{\normalsize{\textbf {Acknowledgements}}}
\vspace{3 mm}
\\
\hspace*{4 mm}We would like to thank the CERN-SPS accelerator crew
for the excellent lead beam provided and 
the Laboratoire National Saturne for the loan of the magnet Goliath.
This work was supported jointly by the German BMBF and DFG, the U.S. 
DOE, the Swedish NFR, the Dutch Stichting FOM, the Swiss National Fund,
the Humboldt Foundation, the Stiftung f\"{u}r deutsch-polnische
Zusammenarbeit, the Department of Atomic Energy, the Department
of Science and Technology and the University Grants Commission of
the Government of India, the Indo-FRG Exchange Programme, the PPE
division of CERN, the INTAS under contract INTAS-97-0158, the
Polish KBN under the grant 2P03B16815, 
the Grant-in-Aid for Scientific Research (Specially Promoted Research 
\& International Scientific Research) of the Ministry of Edication, 
Science, Sports and Culture, JSPS Research Fellowships for Young 
Scientists, the University of Tsukuba Special Research Projects, 
and ORISE.
ORNL is managed by UT-Battele, LLC, for the U.S. Department of Energy
under contract DE-AC05-00OR22725.

\end{document}